\documentclass[twocolumn,twocolappendix,trackchanges,nolinenumbers]{aastex631}
\usepackage{threeparttable}
\usepackage{appendix}
\usepackage{natbib}
\usepackage{color}
\usepackage{mathptmx}
\usepackage{amsmath}
\usepackage{url}
\usepackage{multirow}
\usepackage{verbatim}
\usepackage[utf8]{inputenc}
\usepackage{comment}

\newcommand{\msun}{\ M_\odot}

\definecolor{jhcol}{rgb}{0,0.8,0}

\begin{document}

\title{Line Intensity Mapping Prediction from the Cosmic Dawn (CoDa) III Simulation for H$\alpha$ from Galaxies and the Intergalactic Medium during the Epoch of Reionization}

\author[0000-0002-0786-7307]{Eugene Hyeonmin Lee}
\affiliation{Department of Astronomy and Texas Center for Cosmology and Astroparticle Physics, The University of Texas at Austin, Austin, TX 78712-1083, USA}

\author[0000-0001-8593-8222]{Joohyun Lee}
\affiliation{Department of Astronomy and Texas Center for Cosmology and Astroparticle Physics, The University of Texas at Austin, Austin, TX 78712-1083, USA}

\author[0000-0002-0410-3045]{Paul R. Shapiro}
\affiliation{Department of Astronomy and Texas Center for Cosmology and Astroparticle Physics, The University of Texas at Austin, Austin, TX 78712-1083, USA}

\author{Pierre Ocvirk}
\affiliation{Universit\'e de Strasbourg, Observatoire astronomique de Strasbourg, UMR 7550, F-67000 Strasbourg, France}

\author{Joseph S. W. Lewis}
\affiliation{Institut d’Astrophysique de Paris, UMR 7095, CNRS, UPMC Univ. Paris VI, 98 bis boulevard Arago, 75014 Paris, France}

\author{Taha Dawoodbhoy}
\affiliation{Department of Physics, California Polytechnic State University, San Luis Obispo, California 93402 USA}

\author{Ilian T. Iliev}
\affiliation{Department of Physics \& Astronomy, University of Sussex, Brighton, BN1 9QH, UK}

\author{Luke Conaboy}
\affiliation{School of Physics and Astronomy, University of Nottingham, University Park, Nottingham, NG7 2RD, UK}

\author{Kyungjin Ahn}
\affiliation{Chosun University, 375 Seosuk-dong, Dong-gu, Gwangju 501-759, Korea}

\author{Hyunbae Park}
\affiliation{Center for Computational Sciences, University of Tsukuba, 1-1-1 Tennodai, Tsukuba, Ibaraki 305-8577, Japan}

\author{Jenny G. Sorce}
\affiliation{Univ. Lille, CNRS, Centrale Lille, UMR 9189 CRIStAL, F-59000 Lille, France}

\author{Dominique Aubert}
\affiliation{Universit\'e de Strasbourg, Observatoire astronomique de Strasbourg, UMR 7550, F-67000 Strasbourg, France}

\author{Romain Teyssier}
\affiliation{Department of Astrophysical Sciences, Princeton University, 4 Ivy Lane, 08540, Princeton, NJ, USA}

\author{Gustavo Yepes}
\affiliation{Departamento de Física Teórica M-8, Universidad Autónoma de Madrid, Cantoblanco, 28049 Madrid, Spain}
\affiliation{Centro de Investigación Avanzada en Física Fundamental (CIAFF), Universidad Autónoma de Madrid, 28049 Madrid, Spain}

\author{Yohan Dubois}
\affiliation{Institut d’Astrophysique de Paris, UMR 7095, CNRS, UPMC Univ. Paris VI, 98 bis boulevard Arago, 75014 Paris, France}

\author{Stefan Gottl\"ober}
\affiliation{Leibniz-Institut f{\"u}r Astrophysik Potsdam (AIP), An der Sternwarte 16, D-14482 Potsdam, Germany}




\begin{abstract}

The evolution of large-scale structure, galaxies and the intergalactic medium (IGM) during the Epoch of 
Reionization (EoR) can be probed by upcoming Line Intensity Mapping (LIM)
experiments, which sample in redshift and direction without needing to resolve individual 
galaxies. We predict the intensity and sources of hydrogen H$\alpha$ emission, dominated by radiative recombination following ionization by UV from the same massive stars that 
caused reionization, down to redshift 4.6, using the largest fully-coupled, radiation-hydro simulation of galaxy formation and reionization to date, Cosmic Dawn 
(CoDa) III. We compute the mean intensity and Voxel Intensity Distribution (VID) vs. redshift, including the relative contributions of galaxies and IGM. This will provide mock data to guide and interpret LIM experiments such as NASA’s SPHEREx and proposed Cosmic Dawn Intensity Mapper (CDIM).
\end{abstract}



\section{Introduction} \label{sec:intro}

During the EoR, each H-atom's ionization (minus the last) is followed by a recombination, some fraction releasing an H$\alpha$ photon. Collisional excitation also contributes to this emission. Line Intensity Mapping (LIM) experiments are underway to measure this signal and its anisotropy on angular scales that probe the integrated flux in large spatial volumes that contain multiple sources, rather than resolving individual sources, to probe cosmology, large-scale structure, galaxy formation, and large-scale processes like reionization \citep[see][]{Kovetz}. Additionally, LIM provides statistical information on star formation rate and other galactic properties \citep{Gong17}. While many studies assume this LIM is entirely from galaxies, there are far more H atoms outside galaxies which also recombine following their ionization during the EoR \citep{Shapiro94, Onken04}, so it is important to include this in LIM predictions.

This is a theoretical challenge, since it requires self-consistent radiation-hydro modelling of galaxy formation and reionization that accounts for all sources and sinks of ionizing photons and reionization feedback on galactic SFRs.  We present preliminary results for this, utilizing the Cosmic Dawn (CoDa) III simulation \citep{CoDa3}.

We calculate H$\alpha$ luminosity of each grid cell in CoDa III during the EoR and post-reionization, to redshift 4.6, to construct 3-D line intensity maps and analyze their spatial distribution. We also distinguish the relative contributions from galactic haloes and the intergalactic medium (IGM) by separating them according to dark matter overdensity.  Our goal is to provide LIM predictions, by creating intensity maps at various redshifts, for LIM observation missions such as SPHEREx.

\section{Methods} \label{sec:sec2}
\subsection{Simulation}
CoDa III \citep{CoDa3} is a cosmological simulation of galaxy formation and reionization with fully-coupled radiation-hydrodynamics, using massively-hybrid CPU-GPU code RAMSES-CUDATON \citep{Teyssier02, Aubert08}. 
With a trillion computational elements ($8192^{3}$ dark matter particles and $8192^{3}$ Eulerian grid cells for gas and radiation), in a 94.4 cMpc box, 
CoDa III is the largest such simulation to-date, resolving all galactic halos responsible for reionization ($M>10^8{M}_\odot$). 
We analyze snapshots for 8 redshifts between 15 and 4.64. 

\subsection{H$\alpha$ intensity modeling}
We calculate the H$\alpha$ luminosity of each cell, from spontaneous decays following H recombination and collisional excitation. Recombination luminosity per cell of volume $\Delta V$ is given by:
\begin{equation}
    L_{\alpha}^{\rm rec}=h\nu_\alpha P_{B,\,\alpha} (T,\,n_e)\alpha_B(T) n_e n_p \Delta V,
\end{equation} 
where $h\nu_{\text{H}\alpha}$ = 1.89 eV, 
\( P_B \) is the conversion probability per recombination, \( \alpha_B(T) \) is the case B recombination coefficient at temperature \( T \), and \( n_e \) and \( n_p \) are number densities of free electrons and protons, respectively. 
As fit by \cite{Draine}, 
$P_{B,H_\alpha} \alpha_B$
is approximated by \(\alpha_{\mathrm{eff, H\alpha}}\), given by:
\begin{equation}
\alpha_{\mathrm{eff, H\alpha}} \approx 1.17 \times 10^{-13} T_4^{-0.942 - 0.031\ln T_4} \, \mathrm{cm}^3 \mathrm{s}^{-1}
\end{equation}
where \(T_4\equiv T/10^{4}K\). H$\alpha$ luminosity from collisional excitation is given by:
\begin{equation}
    L_{\alpha}^{\rm col}=h\nu_\alpha q_{{\rm col},\,\alpha}(T) n_e n_{\rm H\,I} \Delta V,
\end{equation}
where \(q_{{\rm col},\,\alpha}(T)\) is the collisional rate coefficient. 
As fit by \cite{Raga2015}, \(h\nu_{\alpha} \times q_{{\rm col},\,X}(T)\) is approximated by:
\begin{equation}
\frac{3.57 \times 10^{-17}}{T^{0.5}} e^{-140360/T} \times \left( 1 + \frac{7.8}{1 + 5 \times 10^{5}/T} \right) \, \mathrm{erg} \, \mathrm{cm}^3 \mathrm{s}^{-1}
\end{equation}

By averaging cell-wise H$\alpha$ luminosities at a given time and dividing by their proper cell volumes, we derived the mean luminosity density $\bar{l}_{\text{H}\alpha}$ vs. time, from which the observed mean intensity $\bar{I}_{\nu_{\text{obs}}}$, is given by \citep{Mao}:

\begin{equation}
    \bar I_{\nu_{\text{obs}}} = \frac{c \bar{l}_{\text{H}\alpha} a \nu_{\text{obs}}^2}{4 \pi \nu_{\alpha}^3 H(a)}
\end{equation}
where 
$\nu_{\rm obs}=a\nu_{\alpha}=\nu_{\alpha}/(1+z)$. 

\begin{figure*}[htbp]
    \centering
    \includegraphics[width=\textwidth]{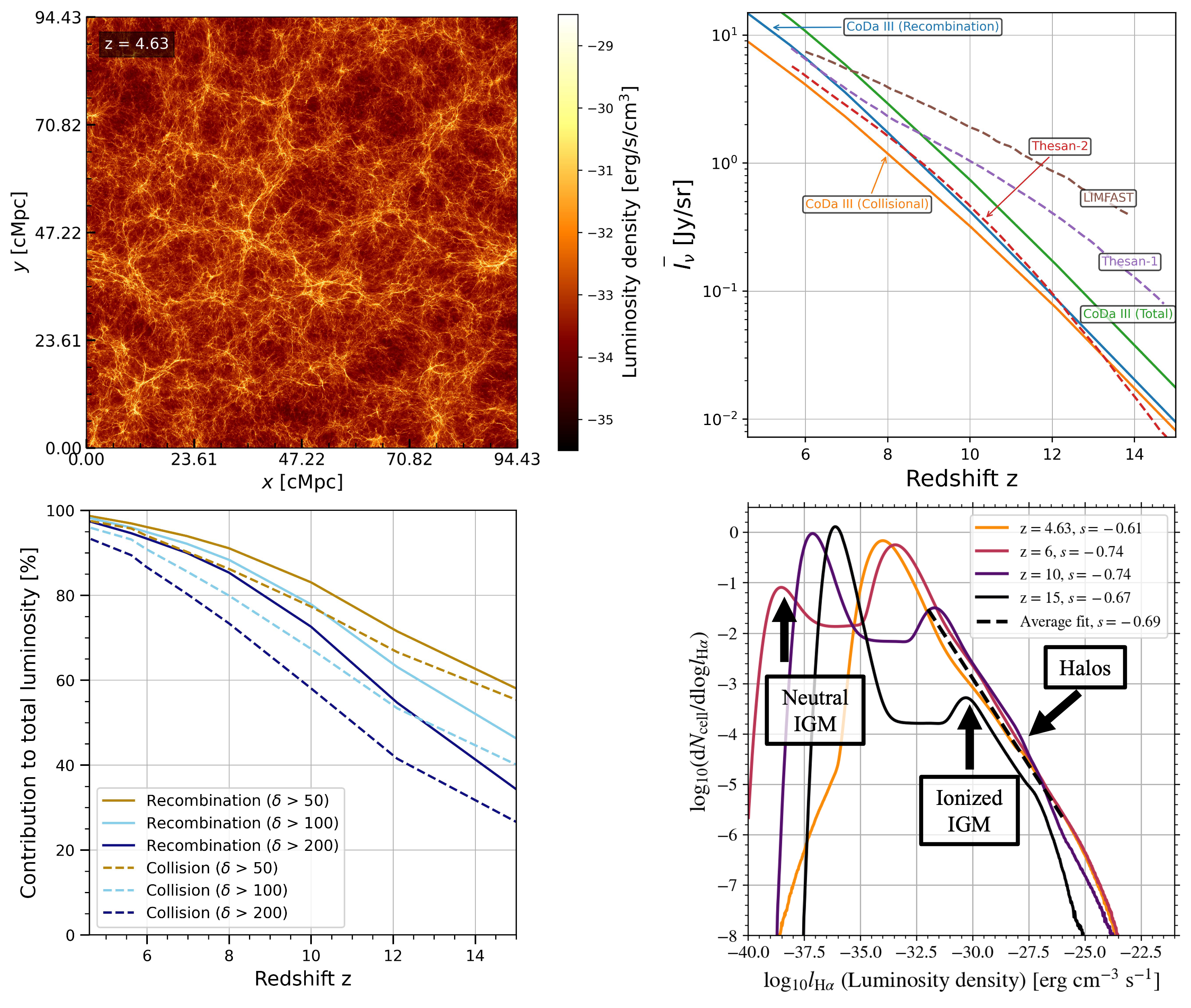}  
    \caption{Upper left: H$\alpha$ luminosity density at post-reionization $z=4.64$ (CoDa III), projected thru 10\% of box. Note: Filamentary structure. 
    Upper right: $\bar{I}_{\rm H\alpha}$ (CoDa III) evolution, compared to {\sc Thesan}-1, {\sc Thesan}-2 \citep{Kannan22} and LIMFAST \citep{Mas_Ribas}. 
    Lower left: Halo (cell overdensity $\delta > 50,\ 100,\,200$) and IGM (cell overdensity $\delta \leq 50,\ 100,\,200$) contributions to H$\alpha$ emission from recombination vs. collisions. 
    Lower right: H$\alpha$ luminosity density PDF of all cells (VID), at different epochs.
    Dashed line indicates the average slope of power-law fit to the bright end at all epochs.
    }
    \label{fig:fig1}
\end{figure*}

\section{Results}

\subsection{H$\alpha$ maps and mean intensity}
Figure \ref{fig:fig1} (upper left) shows a map of luminosity density from CoDa III at redshift 4.6.  
The upper right panel shows $\bar{I}_{\rm H\alpha}$ and compares it with other recent work based on semi-analytical modeling \citep{Mas_Ribas} and simulation \citep{Kannan22}. 
$\bar{I}_{\rm H\alpha}$ increases as redshift decreases. 
At $z=15$, recombinational $\bar{I}_{\rm H\alpha}$ was 53.7\% of the total, increasing to 62.3\% by $z=4.6$.

\subsection{H$\alpha$ emission contributions from IGM vs. halos}

Figure \ref{fig:fig1} (bottom left) shows the changing relative contribution from halos and IGM, if we identify halos with regions of dark matter overdensity $\delta \equiv \rho_{\rm DM} / \bar{\rho}_{\rm DM}$, where $\rho_{\rm DM}$ is the dark matter density of a cell,
with $\delta >50$, 100, or 200, respectively.
The contribution from overdense halo regions increases as redshift decreases, while gas in the IGM and outside dense galactic halo regions grows more important at higher redshift.

\subsection{Voxel intensity distribution}

Voxel intensity distribution (VID) is the probability distribution function of intensity per voxel, a one-point statistic useful in LIM research \citep[e.g.,][]{Sabla2024}, since two-point statistics like power spectra miss a significant amount of information in non-Gaussian fields \citep{LIMPDF}. 
We calculate the VID of our cell-wise H$\alpha$ line luminosity densities in CoDa III, using all $8192^3$ gas cells.

As shown in Figure \ref{fig:fig1} (bottom right), there is a universal slope at the bright end, over orders of magnitude of $l_{\rm H\alpha}$, and a plateau at fainter $l_{\rm H\alpha}$, absent at post-reionization redshift, bounded by two peaks, at high $l_{\rm H\alpha}$ from ionized IGM and low 
$l_{\rm H\alpha}$ from neutral IGM (with epoch-of-recombination``freeze-out''
electron density).  
Both peaks move to lower $l_{\rm H\alpha}$ as
redshift decreases and IGM expands, with ionized-IGM peak amplitude increasing as reionization advances while neutral-IGM peak amplitude decreases, finally disappearing after reionization overlap. 

At the bright end, $dN_{\rm cell}/d\text{log}l_{\text{H}\alpha}$ follows a power-law with slope $s \simeq -0.7$ for $10^{-30} \lesssim l_{\text{H}\alpha}\ [{\rm erg\;cm^{-3}\;s^{-1}}] \lesssim 10^{-27}$.  
We can explain this if halos are responsible, 
with 1) halo mass function in the low-mass regime: $M_{\rm halo}dN_{\rm halo}/dM_{\rm halo} \propto M_{\rm halo}^{\chi_1}$;
2) SFRs $\dot{M_\star} \propto M_{\rm halo}^{\chi_2}$;
3) stellar-to-halo mass ratio $M_\star \propto M_{\rm halo}^{\chi_3}$;
and 4) galactic disk size $r_{1/2} \propto M_{\star}^{\chi_4}$,
yielding
\begin{equation}
    dN_{\rm cell}/d\text{log}l_{\text{H}\alpha}
    \propto l_{\text{H}\alpha}^{\frac{3\chi_{3}\chi_{4} + \chi_{1}}{\chi_{2} - 3\chi_{3}\chi_{4}}}
    \equiv (l_{\text{H}\alpha})^{s}.
\end{equation}
With $\chi_1 = -1$ \citep{PS74},
$\chi_2 = 5/3$ \citep{CoDa1, CoDa2}, 
$\chi_3 = 1.8$ \citep{DUSTiER}, and
$\chi_4 = 0$ in $10^{8} \msun \leq M_{\star} \leq 10^{10} \msun$ \citep{Behroozi22, Somerville25},
$s = -0.6$, close to the fitted value, $-0.7$.




\newpage
\section{Discussion}

Using CoDa III to predict the cosmic $\rm H\alpha$ background with resolution down to kpc scales, we find mean intensity $\bar{I}_{\rm H\alpha}$ and the fraction from dense halo regions increase towards lower redshift. 
The VID shows a universal power-law slope at the bright end, and a plateau at fainter luminosities, bounded by peaks from ionized and neutral IGM regions, respectively, which evolve as reionization advances. 
These features trace the rise of the cosmic web, galaxies, stars, and reionization. 


\newpage

\section{Acknowledgments}
We used resources of Oak Ridge OLCF under DOE INCITE award AST031, Texas Advanced Computing Center at University of Texas at Austin under NSF ACCESS grant PHY240056 and NASA grant No. 80NSSC22K1756.

%

\vspace{5mm}






\bibliography{sample631}{}

\begin{thebibliography}{}
\expandafter\ifx\csname natexlab\endcsname\relax\def\natexlab#1{#1}\fi
\providecommand{\url}[1]{\href{#1}{#1}}
\providecommand{\dodoi}[1]{doi:~\href{http://doi.org/#1}{\nolinkurl{#1}}}
\providecommand{\doeprint}[1]{\href{http://ascl.net/#1}{\nolinkurl{http://ascl.net/#1}}}
\providecommand{\doarXiv}[1]{\href{https://arxiv.org/abs/#1}{\nolinkurl{https://arxiv.org/abs/#1}}}

\bibitem[{{Aubert} \& {Teyssier}(2008)}]{Aubert08}
{Aubert}, D., \& {Teyssier}, R. 2008, \mnras, 387, 295, \dodoi{10.1111/j.1365-2966.2008.13223.x}

\bibitem[{{Behroozi} {et~al.}(2022){Behroozi}, {Hearin}, \& {Moster}}]{Behroozi22}
{Behroozi}, P., {Hearin}, A., \& {Moster}, B.~P. 2022, \mnras, 509, 2800, \dodoi{10.1093/mnras/stab3193}

\bibitem[{{Breysse} {et~al.}(2017){Breysse}, {Kovetz}, {Behroozi}, {Dai}, \& {Kamionkowski}}]{LIMPDF}
{Breysse}, P.~C., {Kovetz}, E.~D., {Behroozi}, P.~S., {Dai}, L., \& {Kamionkowski}, M. 2017, \mnras, 467, 2996, \dodoi{10.1093/mnras/stx203}

\bibitem[{{Draine}(2011)}]{Draine}
{Draine}, B.~T. 2011, {Physics of the Interstellar and Intergalactic Medium} (Princeton Univ. Press, Princeton, NJ)

\bibitem[{{Gong} {et~al.}(2017){Gong}, {Cooray}, {Silva}, {Zemcov}, {Feng}, {Santos}, {Dore}, \& {Chen}}]{Gong17}
{Gong}, Y., {Cooray}, A., {Silva}, M.~B., {et~al.} 2017, \apj, 835, 273, \dodoi{10.3847/1538-4357/835/2/273}

\bibitem[{{Kannan} {et~al.}(2022){Kannan}, {Smith}, {Garaldi}, {Shen}, {Vogelsberger}, {Pakmor}, {Springel}, \& {Hernquist}}]{Kannan22}
{Kannan}, R., {Smith}, A., {Garaldi}, E., {et~al.} 2022, \mnras, 514, 3857, \dodoi{10.1093/mnras/stac1557}

\bibitem[{{Kovetz} {et~al.}(2017){Kovetz}, {Viero}, {Lidz}, {Newburgh}, {Rahman}, {Switzer}, {Kamionkowski}, {Aguirre}, {Alvarez}, {Bock}, {Bond}, {Bower}, {Bradford}, {Breysse}, {Bull}, {Chang}, {Cheng}, {Chung}, {Cleary}, {Corray}, {Crites}, {Croft}, {Dor{\'e}}, {Eastwood}, {Ferrara}, {Fonseca}, {Jacobs}, {Keating}, {Lagache}, {Lakhlani}, {Liu}, {Moodley}, {Murray}, {P{\'e}nin}, {Popping}, {Pullen}, {Reichers}, {Saito}, {Saliwanchik}, {Santos}, {Somerville}, {Stacey}, {Stein}, {Villaescusa-Navarro}, {Visbal}, {Weltman}, {Wolz}, \& {Zemcov}}]{Kovetz}
{Kovetz}, E.~D., {Viero}, M.~P., {Lidz}, A., {et~al.} 2017, arXiv e-prints, arXiv:1709.09066, \dodoi{10.48550/arXiv.1709.09066}

\bibitem[{{Lewis} {et~al.}(2023){Lewis}, {Ocvirk}, {Dubois}, {Aubert}, {Chardin}, {Gillet}, \& {Th{\'e}lie}}]{DUSTiER}
{Lewis}, J. S.~W., {Ocvirk}, P., {Dubois}, Y., {et~al.} 2023, \mnras, 519, 5987, \dodoi{10.1093/mnras/stad081}

\bibitem[{{Lewis} {et~al.}(2022){Lewis}, {Ocvirk}, {Sorce}, {Dubois}, {Aubert}, {Conaboy}, {Shapiro}, {Dawoodbhoy}, {Teyssier}, {Yepes}, {Gottl{\"o}ber}, {Rasera}, {Ahn}, {Iliev}, {Park}, \& {Th{\'e}lie}}]{CoDa3}
{Lewis}, J. S.~W., {Ocvirk}, P., {Sorce}, J.~G., {et~al.} 2022, \mnras, 516, 3389, \dodoi{10.1093/mnras/stac2383}

\bibitem[{{Mao} {et~al.}(2012){Mao}, {Shapiro}, {Mellema}, {Iliev}, {Koda}, \& {Ahn}}]{Mao}
{Mao}, Y., {Shapiro}, P.~R., {Mellema}, G., {et~al.} 2012, \mnras, 422, 926, \dodoi{10.1111/j.1365-2966.2012.20471.x}

\bibitem[{{Mas-Ribas} {et~al.}(2023){Mas-Ribas}, {Sun}, {Chang}, {Gonzalez}, \& {Mebane}}]{Mas_Ribas}
{Mas-Ribas}, L., {Sun}, G., {Chang}, T.-C., {Gonzalez}, M.~O., \& {Mebane}, R.~H. 2023, \apj, 950, 39, \dodoi{10.3847/1538-4357/acc9b2}

\bibitem[{{Ocvirk} {et~al.}(2016){Ocvirk}, {Gillet}, {Shapiro}, {Aubert}, {Iliev}, {Teyssier}, {Yepes}, {Choi}, {Sullivan}, {Knebe}, {Gottl{\"o}ber}, {D'Aloisio}, {Park}, {Hoffman}, \& {Stranex}}]{CoDa1}
{Ocvirk}, P., {Gillet}, N., {Shapiro}, P.~R., {et~al.} 2016, \mnras, 463, 1462, \dodoi{10.1093/mnras/stw2036}

\bibitem[{{Ocvirk} {et~al.}(2020){Ocvirk}, {Aubert}, {Sorce}, {Shapiro}, {Deparis}, {Dawoodbhoy}, {Lewis}, {Teyssier}, {Yepes}, {Gottl{\"o}ber}, {Ahn}, {Iliev}, \& {Hoffman}}]{CoDa2}
{Ocvirk}, P., {Aubert}, D., {Sorce}, J.~G., {et~al.} 2020, \mnras, 496, 4087, \dodoi{10.1093/mnras/staa1266}

\bibitem[{{Onken} \& {Miralda-Escud{\'e}}(2004)}]{Onken04}
{Onken}, C.~A., \& {Miralda-Escud{\'e}}, J. 2004, \apj, 610, 1, \dodoi{10.1086/421378}

\bibitem[{{Press} \& {Schechter}(1974)}]{PS74}
{Press}, W.~H., \& {Schechter}, P. 1974, \apj, 187, 425, \dodoi{10.1086/152650}

\bibitem[{{Raga} {et~al.}(2015){Raga}, {Castellanos-Ram{\'\i}rez}, {Esquivel}, {Rodr{\'\i}guez-Gonz{\'a}lez}, \& {Vel{\'a}zquez}}]{Raga2015}
{Raga}, A.~C., {Castellanos-Ram{\'\i}rez}, A., {Esquivel}, A., {Rodr{\'\i}guez-Gonz{\'a}lez}, A., \& {Vel{\'a}zquez}, P.~F. 2015, \rmxaa, 51, 231.
\newblock \url{https://ui.adsabs.harvard.edu/abs/2015RMxAA..51..231R}

\bibitem[{{Sabla} {et~al.}(2024){Sabla}, {Bernal}, {Sato-Polito}, \& {Kamionkowski}}]{Sabla2024}
{Sabla}, V.~I., {Bernal}, J.~L., {Sato-Polito}, G., \& {Kamionkowski}, M. 2024, \prd, 110, 023507, \dodoi{10.1103/PhysRevD.110.023507}

\bibitem[{{Shapiro} {et~al.}(1994){Shapiro}, {Giroux}, \& {Babul}}]{Shapiro94}
{Shapiro}, P.~R., {Giroux}, M.~L., \& {Babul}, A. 1994, \apj, 427, 25, \dodoi{10.1086/174120}

\bibitem[{{Somerville} {et~al.}(2025){Somerville}, {Gabrielpillai}, {Hadzhiyska}, \& {Genel}}]{Somerville25}
{Somerville}, R.~S., {Gabrielpillai}, A., {Hadzhiyska}, B., \& {Genel}, S. 2025, arXiv e-prints, arXiv:2502.03679, \dodoi{10.48550/arXiv.2502.03679}

\bibitem[{{Teyssier}(2002)}]{Teyssier02}
{Teyssier}, R. 2002, \aap, 385, 337, \dodoi{10.1051/0004-6361:20011817}

\end{thebibliography}
\bibliographystyle{aasjournal}

\end{document}